\documentclass[aps,prl,10pt,preprint,nofootinbib]{revtex4-1}
\usepackage{amsmath}
\usepackage{amssymb}
\usepackage{color}
\usepackage{ulem}
\usepackage{gensymb}
\usepackage{textcomp}
\usepackage{graphicx}
\usepackage{epstopdf}

\usepackage{}

\begin{document}
	
\title{Fingerprint of the Interbond Electron Hopping in Second-order Harmonic Generation}

\author{Liang Li$^{1}$, Tengfei Huang$^{1}$}
\author{Pengfei Lan$^{1}$}\email{Corresponding author: pengfeilan@hust.edu.cn}
\author{Jiapeng Li$^{1}$, Yinfu Zhang$^{1}$, Xiaosong Zhu$^{1}$, Lixin He$^{1}$, Wei Cao$^{1}$}
\author{Peixiang Lu$^{1,2,3}$}\email{Corresponding author: lupeixiang@hust.edu.cn}
		
\affiliation{
$^1$Wuhan National Laboratory for Optoelectronics and School of Physics, Huazhong University of Science and Technology, Wuhan 430074, China\\
$^2$Hubei Key Laboratory of Optical Information and Pattern Recognition, Wuhan Institute of Technology, Wuhan 430205, China\\
$^3$CAS Center for Excellence in Ultra-intense Laser Science, Shanghai 201800, China
}

\begin{abstract}
We experimentally explore the fingerprint of the microscopic electron dynamics in second-order harmonic generation (SHG). It is shown that the interbond electron hopping induces a novel source of nonlinear polarization and plays an important role even when the driving laser intensity is two orders of magnitude lower than the characteristic atomic field. Our model predicts distinct anisotropic structures of the SHG yield contributed by the interbond electron hopping, which is identified in our experiments with ZnO crystals. Moreover, a generalized second-order susceptibility with an explicit form is proposed, which provides a unified description in both the perturbative and strong field regimes. Our work reveals the nonlinear responses of materials at the electron scale and extends the perturbation nonlinear optics to a previously unexplored regime, where the nonlinearity related to the interbond electron hopping becomes dominant. It paves the way for realizing controllable nonlinearity on ultrafast time scale.
\end{abstract}

\maketitle

Nonlinear polarization of materials in laser fields is the basis for manipulating electronic and optical properties with light. Since the observation of second harmonic generation (SHG) in 1961 \cite{Franken}, interest in nonlinear optics has grown continuously, ranging from fundamental studies to applications such as laser frequency conversion \cite{Foster2006,Ettabib2013,Ros2017}, nonlinear microscopy \cite{Tisdale2010,ShenYR2016,ShenYR2004}, optical switching \cite{Nova2019,Fausti2011,McIver2020}, etc. The nonlinear responses are usually explained by perturbation theory within a criterion that the field strength $E$ is much lower than the characteristic atomic field $E_{at}=m^{1/2}(2I_p)^{3/2}/e\hbar$ \cite{Bloembergen,Bloembergen2,Boyd,Shen}, where $e$ and $m$ are the electron charge and mass, and $I_p$ is the ionization potential. Within this framework, the response of a material is linked to a constant nonlinear susceptibility tensor, which is intrinsically determined by the structure of the material. This has inspired many applications for identifying the structural information using nonlinear spectroscopy \cite{Nova2019,ChuanshanT2016,ShenYR2005}. However, how the nonlinear polarization is formed at the electron scale has not been traced \cite{Hassan}.

With the development of laser technology, the laser intensity can be comparable to the Coulomb feld between the valence electron and nucleus ($E\sim E_{at}$). Subject to such a strong field, high-order harmonic generation (HHG) has been observed in a wide range of materials including gases \cite{gas1,gas2}, solids \cite{GhimireS,Jurgens,Luu,Schubert,Ndabashimiye,You2016,liliang2021}, and liquids \cite{TTL2018}. HHG enables the production of attosecond pulse and creates revolutionary changes in time-resolved metrology. Ultrafast time resolution in the few-femtosecond or attosecond regime has provided direct insight into the impulsive responses of materials to electromagnetic fields at the electron scale \cite{Schultze,Lenzner,Ghimire,Zheltikov2014,Zheltikov2020,Durach,Hassan,Vampa,Schiffrin,Higuchi}.
Using high-order harmonic spectroscopy, it has been confirmed that the charge migration and dynamic core polarization play an indispensable role in forming a macroscopic nonlinear polarization, and their features are encoded in the high harmonic spectra \cite{Calegari2014,Kraus2015,Smirnova2019,LL2021}. In contrast to the high order nonlinear polarization formed by the positive-energy electrons, the low order nonlinear response is dominant by the quantum motion of bound electrons and generally has an efficiency of several orders higher than HHG, which therefore suggests more important applications in photonic technology. Yet, how the instantaneous microscopic electron motion is imprinted on the low order nonlinearities has not been well addressed.

In this Letter, we identify the fingerprint of the interbond electron hopping in SHG. Unlike the established perturbation nonlinear optics theory \cite{Bloembergen,Bloembergen2,Boyd,Shen}, we show that the interbond electron hopping induces a novel and unexplored nonperturbative polarization beyond those of the perturbative responses with the form of power series expansion. We propose a non-perturbation bond-charge model (NPBCM) and a generalized second-order susceptibility (GSOS) to model this effect. The proof-of-principle experiment with ZnO crystal indicates that the SHG relevant to the interbond electron hopping shows a distinct anisotropic structure and increases exponentially with the laser intensity. Moreover, the interbond electron hopping becomes dominant even when the driving laser intensity ($\sim$2 TW/cm$^2$) is two orders of magnitude lower than the characteristic atomic field ($\sim$503 TW/cm$^2$ for ZnO).

\begin{figure}[!t]
	\centering
	\includegraphics[width=14cm]{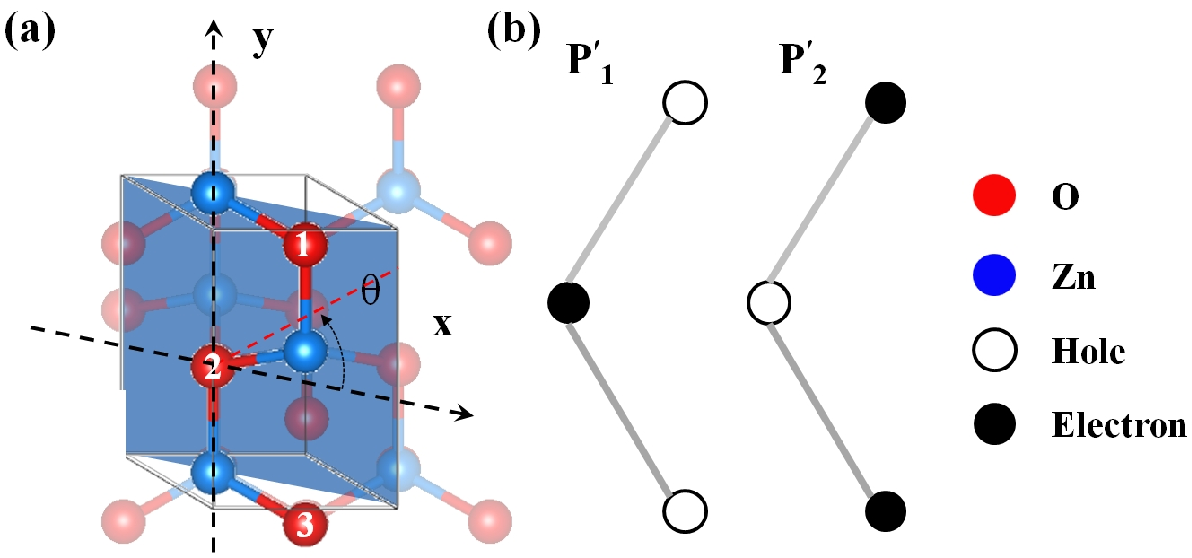}
	\caption{(a) The structure of a ZnO crystal. The x-y plane is highlighted. (b) The sketch of the second-order nonlinear polarization contributed by the interbond electron hopping.}\label{fig1}
\end{figure}

To obtain an intuitive picture, we describe the nonlinear polarization following the well-known bond-charge model (BCM) \cite{Boyd,BCM1,BCM2}. Using the BCM, one can intuitively show the symmetry of the crystal and accurately construct the form of second-order susceptibility accordingly. We adopt ZnO as a prototype to discuss the SHG. In a unit cell of the ZnO crystal, there are eight Zn-O bonds forming two tetrahedral structures (see Fig. \ref{fig1}(a)). We consider that the laser pulse is polarized in the x-y plane, i.e., $(11\overline{2}0)$. By summing the responses of all eight bonds and taking account of the orientation of each bond, one can obtain the second-order nonlinear polarization $\textbf{P}^{0}$ and the second-order susceptibility  $\chi^{(2)}$ (see Sec. A in the supplementary information \cite{SM})

\begin{align}\label{susceptibility_0}
	\left[
	\begin{array}{ccc}
		P_{x}^0\\
		P_{y}^0\\
	\end{array}
	\right]
	=\chi^{(2)}\textbf{EE}
	= \beta^{0}\left[
	\begin{array}{ccc}
		0 & 0 & a\\
		a & b & 0\\
	\end{array}
	\right]
	\left[
	\begin{array}{ccc}
		E_{x}E_{x}\\
		E_{y}E_{y}\\
		2E_{x}E_{y}\\
	\end{array}
	\right]
\end{align}
$\beta^{0}$ is the hyperpolarizabilities of the Zn-O bond. Note that this form of second-order susceptibility matrix can also be derived from the mirror symmetry of the system and the parameters $a$ and $b$ can be determined to $a = -0.8715$ and $b = 1.7760$ by assuming all the Zn-O bonds are equivalent.

In BCM, the outer electrons are assumed localized within the inherent bonds and the coupling between the nearby bonds is neglected. This assumption works well in the weak field regime. However, with increasing the laser intensity, electrons within the bonds can be excited and become less localized. Then, the electron hopping between the nearby bonds will be significant when the electrons become energetic. In this context, we extend the BCM by introducing the interbond electron hopping, which is a nonperturbative effect and thus we call it NPBCM. For ZnO, the upmost valence electrons are primarily from the 2p orbital of oxygen atoms. One can identify that the hopping between the nearest oxygens is dominant, because the barriers between the next-neighbor oxygens are wider and the hopping probability is much smaller. There are mainly two kinds of hopping in one unit cell of the ZnO crystal. The first one is the hopping between oxygen 1 and oxygen 2 (refer to O1, O2). The second occurs between O2 and O3. This process is similar to the charge migration in molecules \cite{migrationm1,migrationm2}. The electron migration between the nearby oxygens forms an additional polarization beyond the inherent Zn-O bonds and contributes a new source of nonlinear polarization. We denote these new nonlinear polarization as $\mathbf{p}_{1}^{\prime}$ and $\mathbf{p}_{2}^{\prime}$ (see Fig. \ref{fig1}(b)). The interbond electron hopping depends sensitively on the driving laser, and therefore, their hyperpolarizabilities are laser-dependent too, which are denoted as $\beta_{1}(E_{0},\theta)$ and $\beta_{2}(E_{0},\theta)$. Finally, one can write the second-order nonlinear polarization by using the susceptibilities, $\textbf{P}^{'} = \textbf{p}_{1}^{\prime}+\textbf{p}_{2}^{\prime} =\chi'^{(2)}\textbf{E}\textbf{E}$, that is (see Sec. B in the supplementary information \cite{SM}),
\begin{align}\label{susceptibility'}
	\left[
	\begin{array}{ccc}
		P_{x}^{\prime}\\
		P_{y}^{\prime}\\
	\end{array}
	\right]
	=\chi^{^{\prime}(2)}\textbf{EE}
	= \left[
	\begin{array}{ccc}
		2\cos^{3}(\theta_{0})\beta^{-} & 2\cos(\theta_{0})\sin^{2}(\theta_{0})\beta^{-} & 2\cos^{2}(\theta_{0})\sin(\theta_{0})\beta^{+}\\
		0 & 0 & 0\\
	\end{array}
	\right]
	\left[
	\begin{array}{ccc}
		E_{x}E_{x}\\
		E_{y}E_{y}\\
		2E_{x}E_{y}\\
	\end{array}
	\right]
\end{align}
where $\beta^{-} = \beta_{2}(E_{0},\theta)-\beta_{1}(E_{0},\theta)$, and $\beta^{+} = \beta_{2}(E_{0},\theta)+\beta_{1}(E_{0},\theta)$. Note that this nonlinear dipole $\textbf{P}^{\prime}$ is formed by the fluctuation of electron densities due to the electron hopping between the nearby oxygens. Thus, the hopping angle $\theta_{0}\simeq0.3\pi$ and the contribution of the interbond electron hopping becomes more dominant when the transition population is higher. In contrast, the inherent bond term will be decreased due to the loss of bond charges. We denote the population of the excited electron as $\rho_e$. Then the total second-order nonlinear polarization is expressed as $\textbf{P} = (1-\rho_{e})\textbf{P}^{0}+\rho_{e}\textbf{P}^{\prime}$ and the GSOS can be defined,

\begin{equation}\label{GSOS}
	\eta^{(2)}_{ijk}(\beta^{0},\beta^{+},\beta^{-})= (1-\rho_e) \chi^{(2)}_{ijk}(\beta^{0})+ \rho_e \chi^{\prime(2)}_{ijk}(\beta^{+},\beta^{-})
\end{equation}

According to Eqs. (\ref{susceptibility_0}), (\ref{susceptibility'}) and (\ref{GSOS}), we can rewrite the GSOS as
\begin{align}\label{GSOS_num}
	\eta^{(2)} &= k_{0}\left[
	\begin{array}{ccc}
		0 & 0 & a\\
		a & b & 0\\
	\end{array}
	\right]
	+\left[
	\begin{array}{ccc}
		2\cos^{3}(\theta_{0})k^{-} & 2\cos(\theta_{0})\sin^{2}(\theta_{0})k^{-} & 2\cos^{2}(\theta_{0})\sin(\theta_{0})k^{+}\\
		0 & 0 & 0\\
	\end{array}
	\right]
\end{align}
where $k_{0} = (1-\rho_{e})\beta^{0}$, and $k^{\pm} = \rho_{e}\beta^{\pm}$. Here, we introduce two parameters $k$ and $\sigma$ to describe the interbond electron hopping terms $\chi^{\prime(2)}(\beta^{+},\beta^{-})$. $k^{\pm} = \rho_{e}\beta^{\pm} = k[\frac{\alpha_{1}}{\max(\alpha_{1})}\pm\frac{\alpha_{2}}{\max(\alpha_{2})}]$, where $\alpha_{1,2} = e^{\sigma(E_{0})|\cos(\theta-\theta_{1,2})|^{2}}$ (see Sec. B in the supplementary information \cite{SM}). $\theta_{1}$ ($\theta_{2}$) marks the angle along the direction from O1 (O2) to O2 (O3). The parameter $k$ describes the weight, and the parameters $\frac{\alpha_{1,2}}{\max(\alpha_{1,2})}$ describe the normalized orientation dependence of the interbond electron hopping terms in SHG.

\begin{figure}[!t]
	\centering
	\includegraphics[width=14cm]{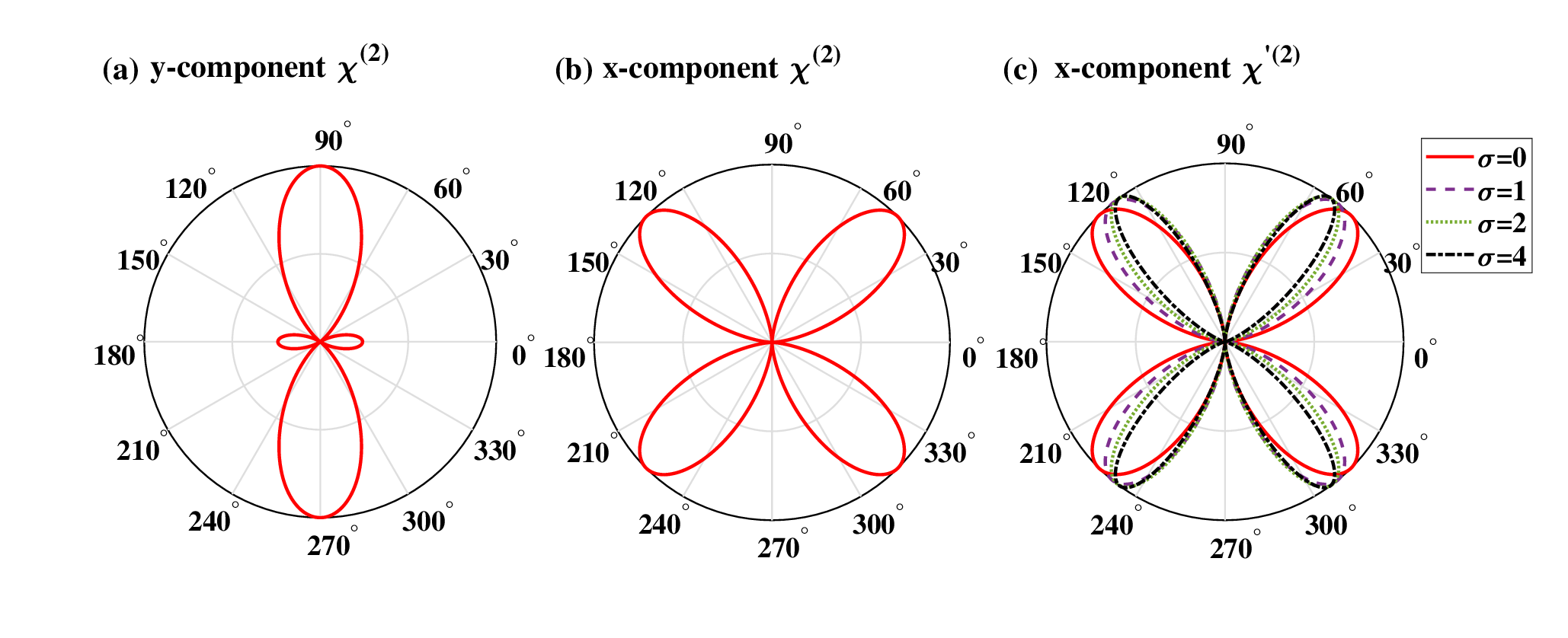}
	\caption{The theoretical results of the anisotropic SHG yields calculated by Eqs. (\ref{susceptibility_0}) and (\ref{susceptibility'}). (a) The y-component and (b) x-component SHG contributed by the inherent term $\chi^{(2)}(\beta^{0})$. (c) The x-component SHG contributed by the interbond electron hopping term $\chi'^{(2)}(\beta^{+},\beta^{-})$ for different parameter $\sigma$. The results are all normalized to 1 for clarity.}\label{fig2}
\end{figure}

According to Eq. \ref{GSOS_num}, one can see that the y-component SHG is only contributed by the inherent bond term $\chi^{(2)}(\beta^{0})$, and its anisotropic structure is independent on the laser parameter (see Fig. \ref{fig2}(a)). On the other hand, the x-component SHG can be contributed by both the inherent bond term $\chi^{(2)}(\beta^{0})$ and the interbond electron hopping term $\chi'^{(2)}(\beta^{+},\beta^{-})$. For $\chi^{(2)}(\beta^{0})$, there is only one nonzero component $\chi^{(2)}_{xxy}$ contributing to the x-component SHG, and its structure is determined by $\left|2E_{x}E_{y}\right|^{2}$. As shown in Fig. \ref{fig2}(b), it has a butterfly shape with peaks at 45$^{\degree}$, 135$^{\degree}$, etc. Different from that, the contribution from the interbond electron hopping term $\chi'^{(2)}(\beta^{+},\beta^{-})$ depends on the laser parameters. With increasing the parameter $\sigma$, which is larger for stronger laser fields, the anisotropic structure shows a butterfly shape, however, the peaks move from 45$^{\degree}$, 135$^{\degree}$ to 60$^{\degree}$, 120$^{\degree}$, etc (see Fig. \ref{fig2}(c)). Therefore, the interbond electron hopping term will induce the change of anisotropic structure of x-component SHG.

To demonstrate and identify the role of the interbond electron hopping in SHG, we perform the experiment with a near-infrared (800 nm or 2300 nm) linearly polarized laser pulse. The experiment is carried out based on a Ti:sapphire laser, which delivers 35-fs, 800-nm pulses at a repetition rate of 1 kHz, with the maximum pulse energy of $7$ mJ. The signal and idler infrared pulses are generated with an optical parametric amplifier (TOPAS-Prime-Plus, Coherent) pumped by the Ti:sapphire laser. The maximum pump energy is $5$ mJ and the maximum output energy is about $600 \rm \mu$J for the signal and $500 \rm \mu$J for the idle pulses. The wavelength of the signal and idler pulses can be changed from 1200 nm to 2600 nm. The polarization direction and intensity of the laser pulses are controlled by the half-wave plate and wire-grid polarizer. The laser beam is normally focused on the $350$-$\rm \mu$m-thick, $(11\overline{2}0)$ ZnO crystal (see Fig. \ref{fig1}(a)), and the SHG is measured with the spectrometer by rotating the ZnO crystal. The spot size at the focal point is estimated to be $50 \rm \mu$m, and $125 \rm \mu$m for the $2300$-nm and $800$-nm lasers, respectively, by using the knife-edge measurement. The polarization of the second harmonic is analyzed by using a wire-grid polarizer. The second harmonic of the $2300$-nm pulse is spectrally resolved by a fiber grating spectrometer (NIRQuest 512 and USB4000, Ocean Optics), and that of the $800$-nm pulse is resolved by a free-space coupled spectrometer (SpectraPro HRS-300, Princeton Instruments) and a CCD camera (PI-MAX4, Princeton Instruments).

\begin{figure}[!t]
	\centering
	\includegraphics[width=14cm]{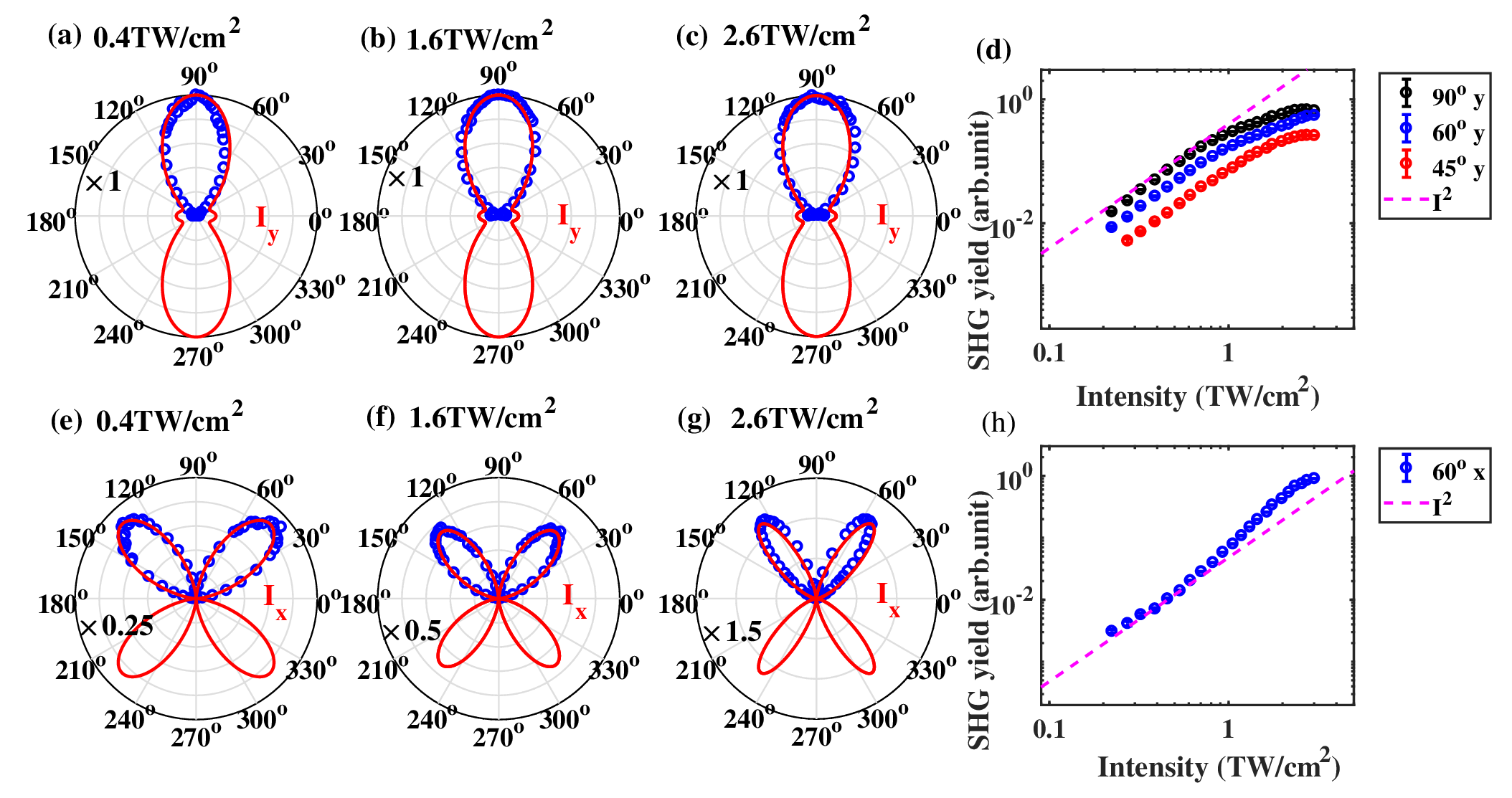}
	\caption{The y-component SHG under the laser intensities (a) 0.4 TW/cm$^2$, (b) 1.6 TW/cm$^2$, and (c) 2.6 TW/cm$^2$. (d) Laser intensity dependence of the y-component SHG yield at representative orientations 45$^{\degree}$, 60$^{\degree}$, and 90$^{\degree}$. The x-component SHG under the laser intensities (e) 0.4 TW/cm$^2$, (f) 1.6 TW/cm$^2$, and (g) 2.6 TW/cm$^2$. (h) Laser intensity dependence of the x-component SHG yield at an representative orientation 60$^{\degree}$. The laser wavelength is 2300 nm. The dots are the experimental data and the red lines are obtained according to the GSOS in Eq. (\ref{GSOS_num}).}\label{fig3}
\end{figure}

The first row in Fig. \ref{fig3} shows the orientation dependence of y-component SHG, and the second row shows the x-component SHG. We have confirmed that the SHG along the propagation direction (z-direction) is negligible (see Sec. C in the supplementary information \cite{SM}). As shown in Figs. \ref{fig3}(a)-\ref{fig3}(c), the y-component shows an ``8'' structure, and the structure is almost unchanged with increasing the laser intensity. The scalings of the y-component at four typical orientations are also shown in Fig. \ref{fig3}(d). One can see that the scalings are nearly $I^2$ below $0.8$ TW/cm$^2$, and then become saturated (increasing slower than $I^2$). In contrast, the x-component behaves differently as shown in Figs. \ref{fig3}(e)-\ref{fig3}(h). First, the orientation dependence of the x-component shows a butterfly structure, and the location of the peaks changes from 45$^\degree$ (135$^\degree$) to about 60$^\degree$ (120$^\degree$) with increasing the laser intensity. Second, the scalings are approximate $I^2$ below $0.6$ TW/cm$^2$ and then show up a faster growth than $I^2$ with increasing the laser intensity until the saturation at around 2 TW/cm$^2$. Moreover, the x-component even becomes dominant over the y-component at 2.6 TW/cm$^2$ (see Figs. \ref{fig3}(c) and \ref{fig3}(g)). Note that such an intensity scaling can be reproduced by varying the laser intensity from the low value to high value and vice versa as long as the maximum intensity is lower than the damage threshold. We also repeat the experiment with the ZnO crystals of different thicknesses, and similar results are observed (see Sec. D in the supplementary information \cite{SM}).

We have also measured the properties of the fundamental pulse before and after transmitting the crystal. The transmission rate of the fundamental pulse is almost isotropy even at the intensity of 2.6 TW/cm$^2$, where, in contrast, the SHG yield shows the butterfly structure and changes with the laser intensity. Our experiments also show that the birefringence of the ZnO crystal does not change under the laser intensities concentrated in our work. These results indicate that the change of the fundamental light passing through the crystal will not lead to new anisotropic component of the refractive index and will not lead to the intensity-dependent modifications of the anisotropic SHG yield (see Sec. E in the supplementary information \cite{SM}). To the best of our knowledge, the phenomena in Fig. \ref{fig3} have never been reported in previous experiments and the perturbation nonlinear optics theory does not predict such intensity-dependent modifications of the SHG.

\begin{figure}[!t]
	\centering
	\includegraphics[width=14cm]{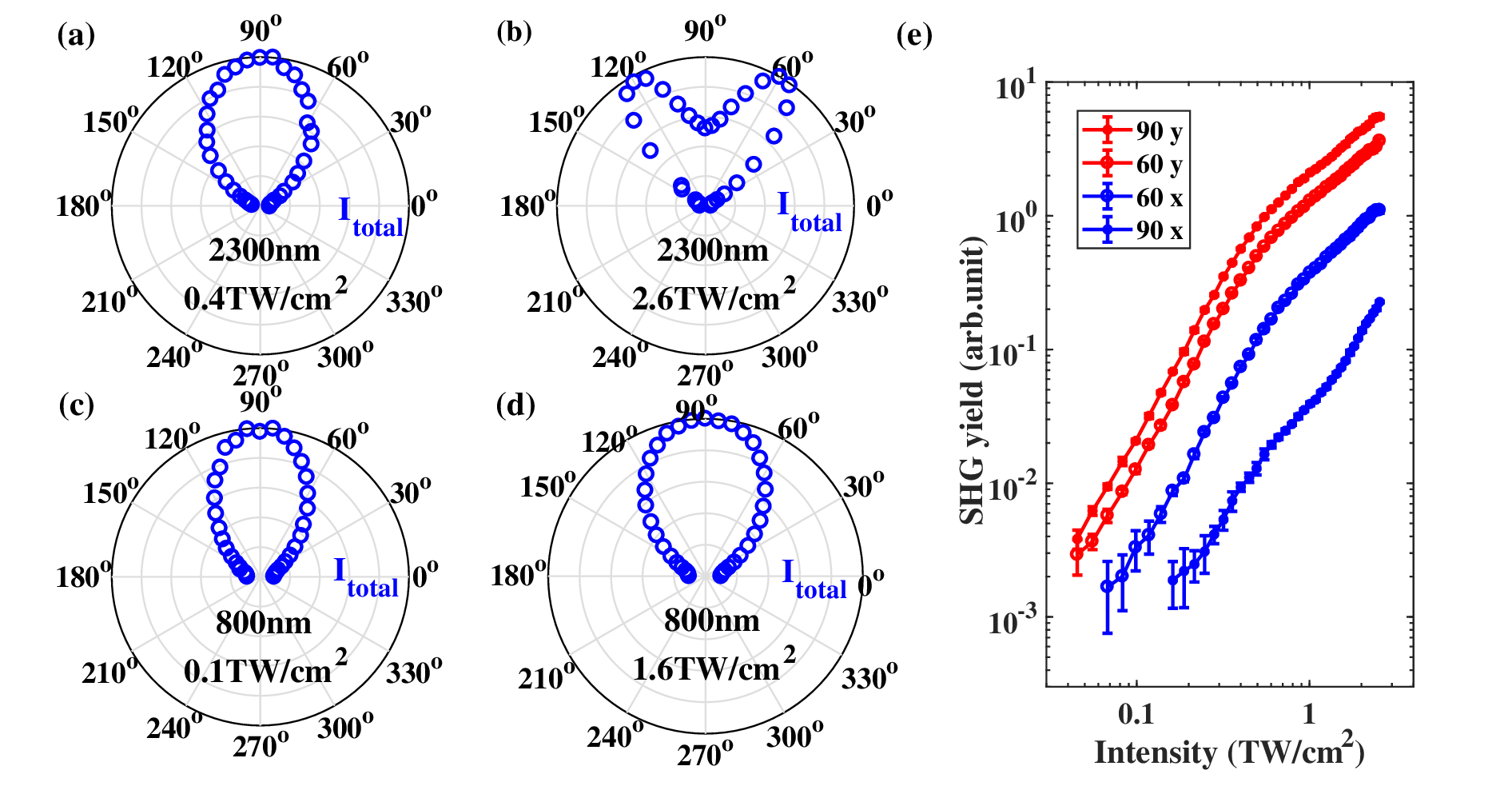}
	\caption{Anisotropic structures of the total SHG yield for the weak and strong laser fields. (a) 0.4 TW/cm$^{2}$ and (b) 2.6 TW/cm$^{2}$ for the 2300-nm laser, and (c) 0.1 TW/cm$^{2}$ and (d) 1.6 TW/cm$^{2}$ for the 800-nm laser. (e) Laser intensity dependence of the x-component and y-component SHG yields at different orientations with the  800-nm laser.}\label{fig4}
\end{figure}

The above experimental results are remarkable fingerprint of the interbond electron hopping effect and can be well explained with our NPBCM. According to Eq. \ref{GSOS_num}, the y-component SHG is only contributed by $\chi^{(2)}(\beta^{0})$. Then one can normalize the y-component SHG, and $k_{0} = 0.56$ can be obtained by fitting the normalized y-component SHG. When the laser field is weak, $\rho_e$ and the coupling between the nearby bonds can be neglected, and the hyperpolarizability $\beta^{0}$ is a laser-independent constant, which goes back to the normal second-order susceptibility. In this case, the SHG yield follows $I^2$ scaling and the anisotropic structures will be unchanged. With increasing the laser intensity, inherent term $\chi^{(2)}(\beta^{0})$ decreases, leading to the saturation structure in Fig. \ref{fig3}(d). These phenomena are in good agreement with the results obtained with GSOS (see red lines). Then, we come to the x-component SHG. By fitting the experimental result of x-component SHG (Figs. \ref{fig3}(e)-(g)), one can obtain the optimal fitting parameters $k =$ 0.02, 0.17, 0.66, and $\sigma = $ 0.52, 3.21, 6.69, for the laser intensities 0.4, 1.6, and 2.6 TW/cm$^{2}$, respectively. The increasing weight parameter k indicates that the interbond electron hopping term becomes more dominant with increasing the laser intensity. According to Eq. \ref{GSOS_num}, the x-component SHG is contributed by both the inherent bond term and the interbond electron hopping term. In weak fields, the SHG is mainly contributed by the inherent bond term and a butterfly structure with peaks at 45$^{\degree}$ and 135$^{\degree}$ (similar to that in Fig. \ref{fig2}(b)) is observed. However, with increasing the laser intensity, both the weight and orientation parameters $k$ and $\sigma$ become larger and the contribution of the interbond electron hopping term becomes dominant. Then, the anisotropic structure of the total SHG yield changes from the ``8'' (see Fig. \ref{fig4}(a)) to the double-peak butterfly (see Fig. \ref{fig4}(b)) shape with increasing the laser intensities. Moreover, the interbond electron hopping term has an approximately exponential growth, which explains the growth faster than $I^{2}$ in Fig. \ref{fig3}(h).

The contribution of the interbond electron hopping term is proportional to the population of excited electrons. The interbond hopping is a two-step process. First, the electron is excited from valence to conduction band. Then, it is accelerated by the laser field and tunnels out through the barrier. Its contribution will be decreased if a shorter wavelength laser is applied because the electron can not be accelerated to high enough energy and the tunnelling is suppressed. To demonstrate this idea, we repeat the above experiments using an 800-nm laser. As shown in Figs. \ref{fig4}(c) and \ref{fig4}(d), the change of the orientation dependencies of SHG becomes less remarkable than that using the 2300-nm laser. Although the relative strength of the x-component SHG is increased with increasing the laser intensity, it is still much weaker than the y-component SHG (see Fig. \ref{fig4}(e)). Therefore, the orientation dependencies of the total SHG only show imperceptible changes until the damage intensity. This result supports our model again and also explains why the phenomena in our experiments were not observed in previous works with the routinely used Ti:Sapphire laser.

In conclusion, we demonstrate the interbond electron hopping as a new source of nonlinear polarization and it will be dominant for SHG even when the driving laser intensity is two orders of magnitude lower than the characteristic atomic field. Our work covers the edge between the perturbation and strong field nonlinear optics. Different from the normal second-order susceptibility, the GSOS involves an explicit laser dependent term. It provides an additional degree of freedom for optical manipulation. One can selectively excite the bond couplings and control their relative strength without modifying the crystal structure. By using few-cycle \cite{Schubert,Schultze} or two-color pulses \cite{Lan,twocolorion}, the nonlinear responses could be tailored on the attosecond time scale, which suggests a novel approach to control the nonlinear responses with high switching rates.

\begin{acknowledgments}
This Letter was supported by the National Key Research and Development Program of China (2017YFE0116600), and National Natural Science Foundation of China (NSFC) (No. 11934006, No. 91950202, No. 11627809, No.12021004, No. 11874165).

L. L. and T. H. contributed equally to this work.
\end{acknowledgments}

\end{document}